\documentclass[fleqn,usenatbib,onecolumn]{mnras}

\usepackage{newtxtext,newtxmath}

\usepackage[T1]{fontenc}
\usepackage{ae,aecompl}

\usepackage{graphicx}	
\usepackage{amsmath}	

\usepackage{amssymb}	

\usepackage[normalem]{ulem}
\usepackage{threeparttable}
\usepackage{ragged2e}
\usepackage{xcolor}
\usepackage{color}

\title{Handy Relation Between Binary Black Hole Merger Times and Host Galaxy Properties}

\author[Holley-Bockelmann, Khan, Williams]
{Kelly Holley-Bockelmann$^{1,2}$\thanks{E-mail: k.holley@vanderbilt.edu},
Fazeel Mahmood Khan$^{3,4}$,
Isaiah Williams$^{1}$, Jaelyn Roth$^{1}$,
\newauthor Michael Rizzo Smith$^{1}$, Kaitlin Porter$^{1}$, Jillian Bellovary$^{5,6,7}$,
Andrea Derdzinski$^{2}$, 
\newauthor Andrea V. Macci\`o$^{3,4}$\\
	$^{1}$Department of Physics and Astronomy, Vanderbilt University, Nashville, TN 37240, USA\\
	$^{2}$Department of Physics, Fisk University, Nashville, TN 37208, USA\\
	$^{3}${New York University Abu Dhabi, P.O. Box 129188, Abu Dhabi, United Arab Emirates}\\ 
    $^{4}${Center for Astrophysics and Space Science (CASS), New York University Abu Dhabi, PO Box 129188, Abu Dhabi, UAE}\\
	$^{5}$ Department of Physics, Queensborough Community College, Bayside, NY, 11364, USA\\
    $^{6}$ Department of Astrophysics, American Museum of Natural History, New York, NY 10024, USA\\
    $^{7}$ Astrophysics Program, CUNY Graduate Center, New York, NY, 10016, USA\\
}
\date{Accepted XXX. Received YYY; in original form ZZZ}

\pubyear{2025}

\begin{document}
	\label{firstpage}
	\pagerange{\pageref{firstpage}--\pageref{lastpage}}
	\maketitle
	
	\begin{abstract}

Over the past 15 years, the evidence has clearly demonstrated that massive black hole (MBH) binary merger timescales depend strongly on the structural and kinematic properties of their host galaxy. Stellar density, gas content, shape and kinematics all play a role, combining in non-linear ways to effect the evolution of the binary. The binary properties themselves, such as eccentricity, mass ratio, and orbital plane, all matter as well. This makes it nontrivial to estimate accurate cosmological MBH binary merger rates, or to generate merger rate ranges that reflect the distribution of galaxy hosts and orbits. Using an extensive set of high-resolution direct $N-$body simulations in which the shape, structure, and kinematics of each galaxy host are directly informed by observations, we map out MBH binary merger timescales over a range of galaxy hosts and MBH binary orbits. This yields a convenient set of scaling relations to determine MBH binary merger timescales -- and the range of merger timescales -- as functions of basic observables. Such scaling relations can be readily employed as a subgrid model in cosmological or semi-analytic studies, for example, to model event rates for LISA or pulsar timing.

	\end{abstract}


	

\section{Introduction}\label{sec-intro}

The low frequency gravitational wave sky is expected to reveal a population of massive black hole (MBH) binaries $( 10^4\text{--}10^9 M_{\odot})$ as they inspiral and coalesce deep within the cores of galaxies throughout the universe. Pulsar timing arrays~\citep{taylor2021nanohertz} may already have detected the most massive part of this population $( \gtrsim 10^8 M_{\odot})$ with the 2023 observation of a stochastic gravitational wave background in the nanohertz frequency band that is consistent with the ensemble of many MBH binaries on decade-long orbits~\citep{NANOGrav:2023, E+IPTA, PPT23, CPTA, Miles25}. While the pulsar timing detection of a nanohertz gravitational wave background is clear, there is some tension between pulsar timing observations and theoretical models because the measured background is louder than the predicted one; if the stochastic background is produced solely by MBH binaries, it may point to a population that is more massive, more numerous, and/or evolves more quickly than expected~\citep{Agazie23_constraints, EPTA2024, Laal25}. With the Laser Interferometer Space Antenna (LISA), an ESA/NASA space-based gravitational wave observatory set to launch in 2035, the merger of $10^4-10^7 M_\odot$ binaries is expected to be detectable out to redshift 20 with signal-to-noise ratios in the hundreds~\citep{redbook}, which will help us piece together a more complete picture of MBH assembly over the whole mass spectrum and through cosmic time. 

Inferring MBH formation and evolution from LISA and pulsar timing observations requires high-fidelity predictions of MBH merger rates and demographics~\citep[e.g.][and references therein]{KHB10,Chen17,Taylor17, Colpi19}. Unfortunately, MBH merger rate predictions span two orders of magnitude~\citep{LISA:2023}, and most of this uncertainty boils down to an uncertainty in the MBH merger timescale~\citep[e.g.][and references therein]{Merritt13}. The reasons for this abound: cosmological simulations and semi-analytic models cannot incorporate the relevant gas, stellar, and gravitational physics over 12 orders of magnitude in spatial scale as the MBHs make their way from separate galaxies to coalescence and therefore each study must choose which physics to approximate.  Broadly, dynamical friction drives the MBHs to the galactic center, where they meet and form a binary. Thereafter, gas drag and stellar scattering shrinks the binary orbit by a factor of $\mathcal{O}( 10^2 -10^4)$ until gravitational wave emission dominates and the binary coalesces \citep{begelman+80}.

A common practice adopted by cosmological simulations is to artificially migrate the MBH towards the center of their host galaxy's potential, essentially equating the MBH merger rate with the galaxy merger rate~\citep[e.g.][]{Sijacki15,Schaye15,Huang18}.  This practice omits the binary's dynamics and associated delays that may result in the process. More sophisticated estimates incorporate dynamical friction on the infalling black hole on-the-fly or through a post-processing analysis~\citep[e.g.][]{Micic11, just11, Antonini+12, Tremmel15, Dosopoulou18, Barausse20, Ma23}. In cosmological simulations, this brings the MBHs to within the spatial resolution of the simulation ($\sim100$ pc); at this separation, the MBHs are not energetically bound as a binary, but are forced to merge nonetheless~\citep[e.g.][]{Tremmel+18, Genina24}. A huge step forward in this field has been made possible recently with the public release of KETJU, which changes the way gravitational interactions between massive black hole particles and collisionless point particles are treated such that gravitational softening is no longer needed for close interactions~\citep{Mannerkoski23}; currently, the expense of the regularization technique prohibits its use on large uniform cosmological volumes, but KETJU's performance in zoom-in simulations and isolated galaxy mergers make it extremely promising for the future~\citep[e.g.][]{Rawlings25, Attard24, Partmann24}.

Explicitly evolving the MBHs through the stellar scattering phase has not been implemented within a cosmological simulation, so \citet{seskha+15} proposed a recipe to estimate the MBH binary lifetime based on a suite of $N-$body simulations and scattering experiments. Their prescription relied on the central stellar density and velocity dispersion at MBH binary influence radius. A similar model was proposed by \citet{vasiliev+15} based on mergers of cuspy galactic nuclei. These prescriptions have been extensively used both in semi-analytic models and in post-processing of cosmological simulations to estimate the MBH binary lifetime in the post-galaxy merger phase \citep{Kelley17, bia19,volonteri2020,Bortolas2021,Sykes2022,Chen2022,NANOGrav:2023,Truant25}.  However, the sphere of influence is unresolved in cosmological simulations, so the key inputs into the recipe are estimated through inward extrapolation of host galaxy profiles at scales hundreds of times larger, yielding inaccurate results. To add to the complication, it has been shown in many recent studies that kinematic and structural properties of the galaxy host such rotation \citep{holley+15,Rasskazov+16,Mirza+17,Avramov21}, bars \citep{bortolas2022} and nuclear star clusters \citep{khan2021,khan2024,muk2025} play a significant role in determining the fate of the massive binary. 

The goal of this paper to find an empirical relation between MBH binary coalescence time and host galaxy properties and orbital configuration. This effort sacrifices deep understanding of the precise dynamical processes at play in a given binary black hole interaction in favor of a global fit of the timescale as a function of relatively easy-to-measure observables from a real or simulated galaxy population.
Here we present a model based on a set of $N-$body simulations of MBH binary evolution in realistic galaxy hosts that contains a range of features -- such as rotation, triaxiality and central nuclear star clusters -- that have been shown to affect MBH coalescence timescales. Our improved prescription is based on parameters like galaxy stellar mass, effective radius, and $v/\sigma$ that are easy to obtain both from cosmological simulations and from observations.

The paper is organized as follows: section 2 presents the features used in the study, including how we constructed and ran the simulation suite; section 3 discusses methods for identifying a robust and predictive relation between coalescence time and galaxy properties; section 4 presents two models that predict coalescence time with a fractional error less than $40\%$ based on observable properties of the host galaxy and black hole orbital configuration; section 5 summarizes, points out caveats and outlines future work.

\section{The Data and Features}
Over the course of a decade, we collected a suite of 28 high-resolution direct $N$-body simulations of MBH binary coalescence within host galaxy models whose structural and kinematic properties are based on observations. These galaxy models span the range of galaxy types, from faint dwarf ellipticals, to classical dwarfs, to the Milky Way, to M87. Many, but not all, of these simulations have been featured in other publications~\citep[e.g.][]{Khan+15,KHB15,khan2021,khan2024}, and typically take months of wallclock time to run on parallel GPU architectures.  Although our sample features multi-component systems with bulges, disks, and often NSCs, we did not include barred or irregular galaxies. Most have clear central black hole measurements, though admittedly none are known to have MBH binaries. Our equilibrium models focus on the galactic center, which, depending on the model, is roughly a few hundred parsecs. Table 1 outlines the galaxy and central MBH parameters. We emphasize that when the galaxy model name is listed, the parameters are set by observational measurements; otherwise the parameters are chosen from a distribution of measurements for that galaxy class. 
For this study, we are interested in isolating the effect of stellar scattering on driving the binary evolution, so the models are all gas- and dark matter-free~\footnote{This close to the galactic center, the dark matter content is a only a few percent of the mass in any event}. We generate an N-body realization using AGAMA~\citep{Vasiliev19}, which allows us to sample the orbits of each potential using the Schwarzschild method, assuming the Jeans approximation to assign velocity components.

Each run is listed in Table 2. In general, we launch a secondary black hole at several times the radius of influence of the primary black hole with an initial eccentricity of 0.5 on either a prograde or retrograde orbit when the underlying galaxy is rotating. At this separation, we are able to observe the last part of the dynamical friction phase as well as the binary pairing and hardening phase, stopping the run once the binary separation is small enough that gravitational radiation would be significant. We integrate the evolution using a 4th-order Hermite integration scheme in $\Phi$-GPU, a GPU-accelerated direct N-body code~\citep{phigpu}. $\Phi$-GPU has optional softening to prevent stellar binary formation; we employ $\epsilon_{\rm BH} = 0$ for BH-BH interactions and $\epsilon_{\rm star}=10^{-2}$ parsec for star-star interactions. For BH-star interactions, the softening parameter is:
\begin{equation}
\epsilon_{\rm{\star,MBH}} = \sqrt{\frac{\epsilon_{\star}^2 + \epsilon_{\rm MBH}^2}{2}}.
\end{equation} \label{eq:soft}

Since our goal is to relate MBH merger timescales to bulk galaxy and orbit properties, the ideal features are those that can be relatively easy to measure with observations or cosmological simulations. We began with the set below. Although many of the features in this list are certainly degenerate, we decided not to preselect the important features for our coalescence-time prediction. In the next section, we describe our treatment of correlations between features in determining our final relation.

\begin{itemize}

\item S{\'e}rsic index, n.  This formally measures the shape of the surface brightness profile~\citep{ser63,ser68}, and can be transformed to a 3-dimensional stellar density profile, $\rho(r)$  using \citet{pru97} (see also \citet{ter05}): 

\begin{equation}
\rho(r) = \rho_0 \left( \frac{r}{R_e} \right)^{-p} e^{-b(r/R_e)^{(1/n)}},
\end{equation} \label{eq:sersic} 

\noindent where $r$ is the radial distance from the galaxy center, n is S{\'e}rsic index, and $R_e$ is effective radius. $\rho_0$ is normalization density that depends on total galaxy mass and effective radius. The power-law index, $p$, and the parameter $b$ can be expressed as: $p = 1.0-0.6097/n + 0.055 63/n^2$,  and $b \approx 2\,n-1/3+0.009876/n$, for $\sim 0.5 < n < 10$  \citep{ter05}.

\item Binary Black Hole Mass, $M_{\rm BBH}$, Primary Black Hole Mass, $M_{\rm BH1}$, Secondary Black Hole Mass, $M_{\rm BH2}$ $[M_\odot]$, and Mass Ratio, $q$.  

\item Galaxy Stellar Mass, $M^*_{\rm gal} \, [M_\odot]$ and Effective Radius, $R_{\rm eff} \, [{\rm pc}]$. Once again, the combination of these features is not completely independent of S{\'e}rsic index above and stellar density in the next section.

\item Nuclear Star Cluster Mass, $M_{\rm NSC} \, [M_\odot]$  and Nuclear Star Cluster Effective Radius, $R_{\rm NSC} \, [{\rm pc}]$. 

\item Ratio of Stellar Rotational Velocity and Velocity Dispersion, $v/\sigma$. We measured this at the influence radius. 

\end{itemize}

We also considered features that could be measured through numerical simulations but would be difficult or impossible to observe. Though this set would be less widely applicable, we included them in case we couldn't predict the binary black hole merger timescale with the above set alone.

\begin{itemize}

\item Orientation of the Binary Black Hole Orbital Plane, ${\rm Cos}\, \Theta$. In our configuration, $\Theta$ is the angle between the angular momentum axis of the binary black hole orbit and the angular momentum axis of the stellar component of the galaxy, and $\Theta=0^{\circ}$ is a planar prograde orientation (corotation), while $\Theta=180^{\circ}$ represents a planar retrograde orientation (counterrotation). Our input suite contains only configurations that are purely co- or counterrotating, but we have a suite (Table 3) that varies ${\rm Cos} \, \Theta$ between -1 and 1, which we use to test the fit.  Note: we combined Cos \,$\Theta$ and $v\over \sigma$ to compactly express galaxy rotation as projected onto the black hole binary orbital plane, $v/\sigma|_\textrm{proj}$.

\item Initial Central Stellar Density, $\rho_{0}$, Stellar Density at the Radius of Influence of the Primary Black Hole, $\rho_{\rm infl}$ $ [M_{\odot}/{\rm pc^3}]$, and Stellar Density at the Radius of Influence of the Binary Black Hole, $\rho_{\rm infl, b}$ Note that these features are highly correlated with the S{\'e}rsic index.

\item Hardening Radius, $r_{h} \, [{\rm pc}]$, and Stellar Density at the Hardening Radius, $\rho_{\rm h}$ $ [M_{\odot}/{\rm pc^3}]$.  The hardening radius is the separation between the black holes after binary formation such that further stellar encounters increase the binary's binding energy, `hardening' the binary. The hardening radius can be expressed by: $r_{ h} = G \, M_2 / 4 \, \sigma_\star^2$, where $M_{\rm BH2}$ is the secondary black hole mass, and $\sigma_\star$ is the stellar velocity dispersion near the binary~\citep{Heggie75}.

\item Initial Eccentricity, $e_{i}$, and Eccentricity at the Hardening Radius $e_{r_h}$. Note that $e_{r_h}$ is correlated to ${\rm Cos}\, \Theta$, in that eccentricity is boosted to radial orbits in counter-rotating orientations ~\citep{Iwasawa+11,Mirza+17}.

\item Hardening Parameter, $s$ $[{\rm pc^{-1}/Myr}]$.  We note that $s$ is a well-known measure of the shrinking rate of a binary black hole orbit and is often taken as a proxy for the merger time.  It is known to be highly correlated to central stellar density ~\citep{quinlan+96}.

\end{itemize}

\begin{figure*}
    \centering{
   \includegraphics[scale=0.5,angle=0]{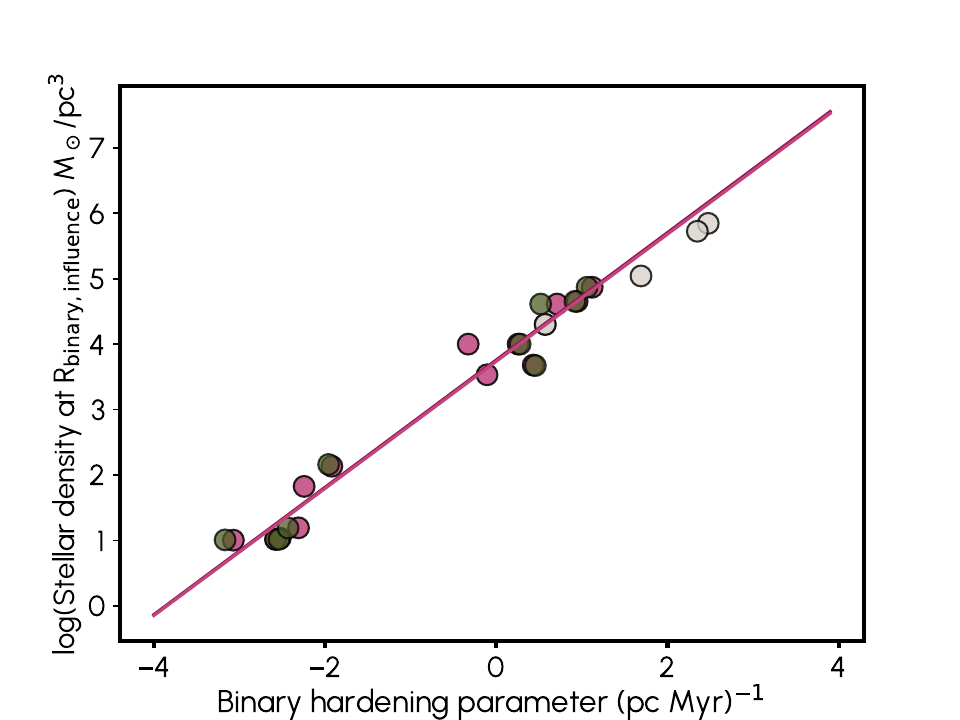}\includegraphics[scale=0.5,angle=0]{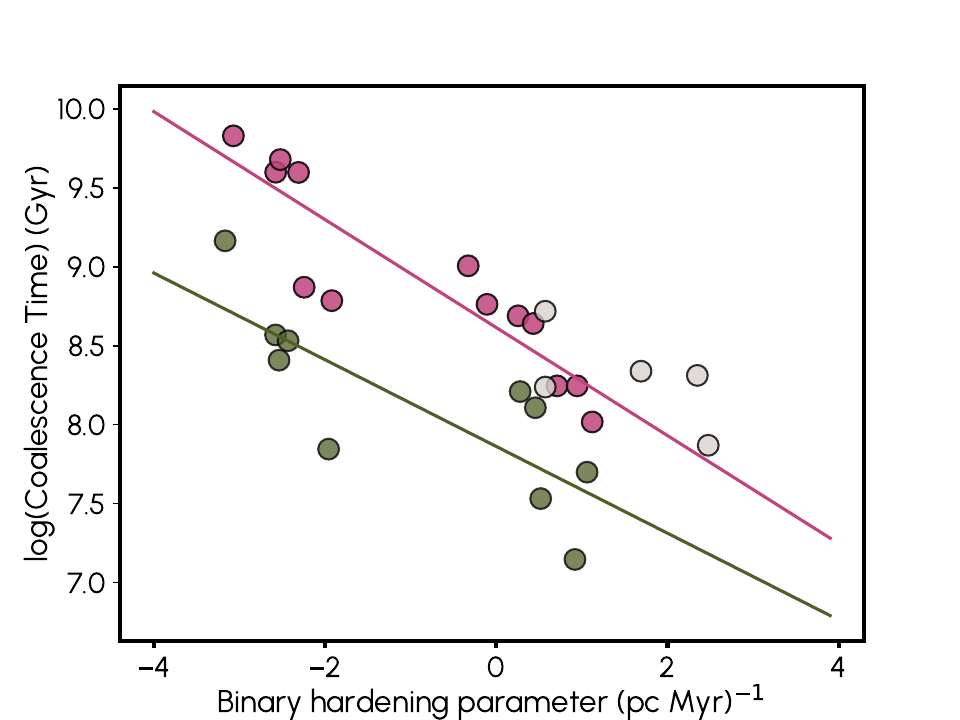}
   }
\caption{Left: Stellar density of the host galaxy at the binary radius of influence, $\rho_{\textrm{infl,b}}$, versus $s$, the binary hardening parameter. Colored dots indicate the orientation of the binary black hole orbital angular momentum with respect to that of the host galaxy. buff:no rotation; pink:retrograde; green:prograde.  Over 6 orders of magnitude in stellar density, there is a tight relation between stellar density and binary black hole hardening, with only minor differences due to the orbit, kinematics, or shape. Note the best fit for the pro- and retrograde are both plotted in pink and green, respectively; they are just indistiguishable on this figure. This indicates that the simulations are properly resolving the few-body scattering phase and is consistent with the finding that the energy extraction from a binary orbit depends on the stellar density in the loss cone~\citep[e.g.][]{quinlan+96}. Right: Binary hardening parameter, $s$, versus coalescence time, log(T$_\textrm{coal})$. For a given hardening parameter, there is between 1-2 orders of magnitude difference in the coalescence time, broadly separated by binary orbit, with prograde configurations taking longer to merge than their retrograde counterparts.}
    \label{fig:hardening}
\end{figure*}

\begin{figure*}
    \centering{
    \includegraphics[scale=0.38,angle=0]{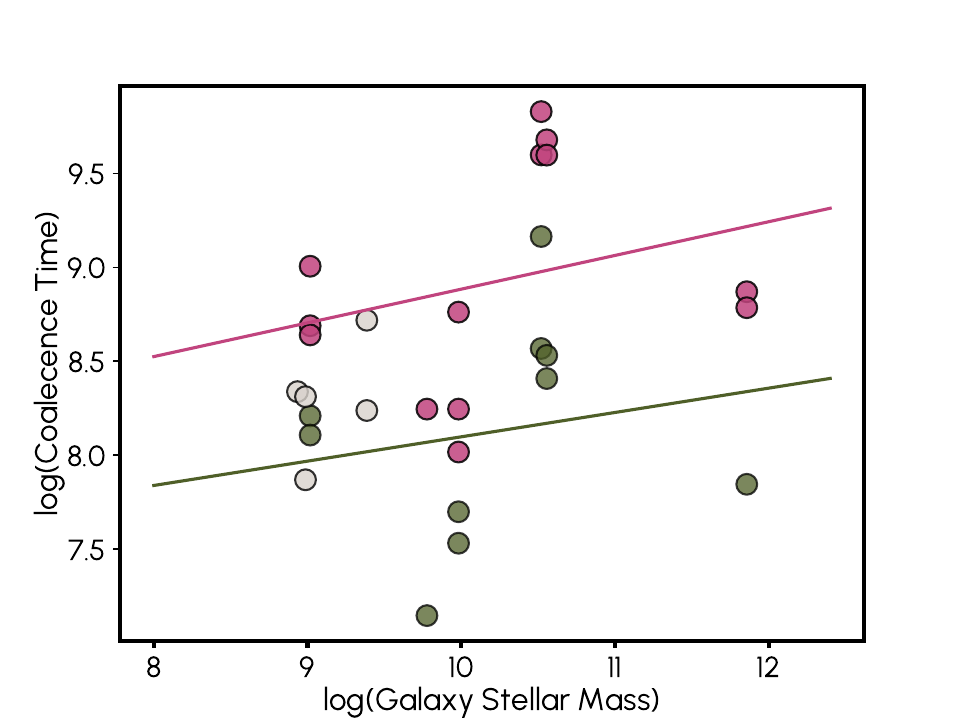}\includegraphics[scale=0.38,angle=0]{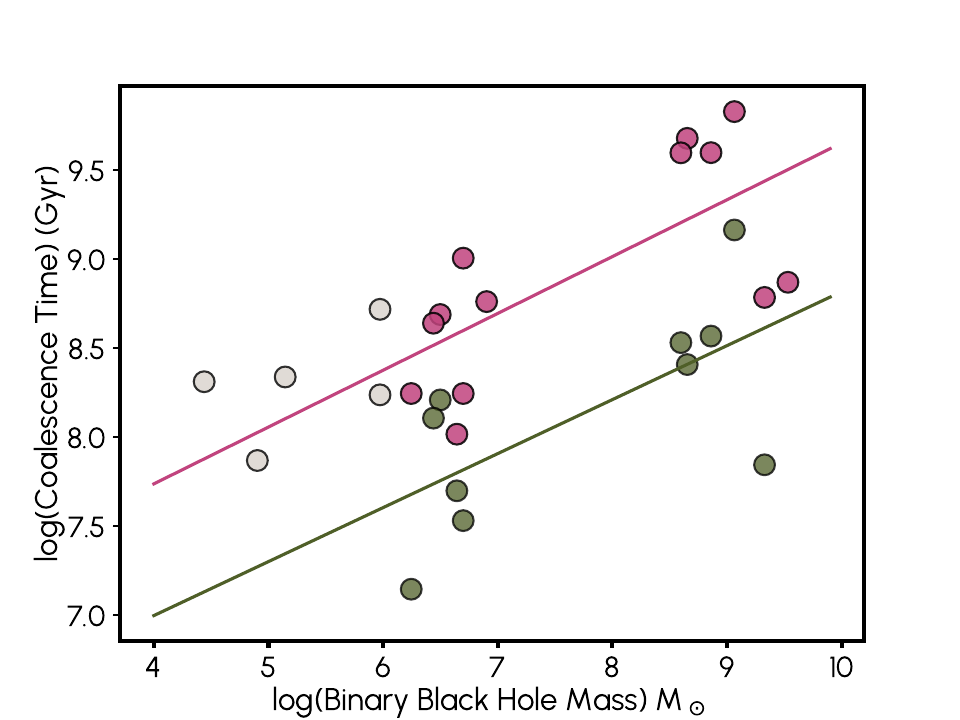}\includegraphics[scale=0.38,angle=0]{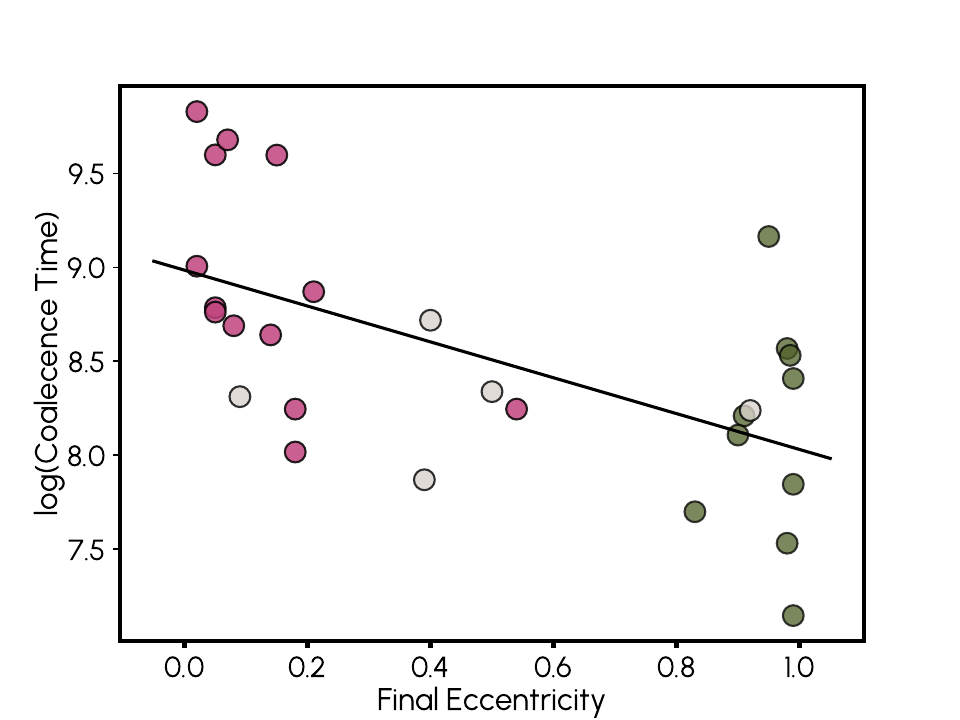}
    }
    \caption{MBH binary coalescence time as a function of selected features. In each panel, the dots are colored according to the orientation of the binary orbital angular momentum vector with respect to that of the host galaxy. green:prograde; buff:non-rotating host; pink:retrograde orbits. Left: We select a host galaxy-related feature, the galaxy stellar mass. Middle: Here, we feature coalescence time as a function of a black hole-related feature, the binary black hole mass, M$_\textrm{BBH}$. Right: We show how coalescence time depends on the eccentricity in the hardening radius, $e_{r_h}$.  Note the large degree of scatter in the coalescence time in each panel, as well as the strong clustering of coalescence time with binary orbit; these are indications that these features are a projection of a higher-dimension manifold, and that the orbit is one of the key predictors of coalescence time.}
    \label{fig:eccentricity}
\end{figure*}

To better illustrate the problem and as a check on the accuracy of our simulation suite, the left panel of Figure~\ref{fig:hardening} displays the hardening parameter, s, as a function of the stellar density at the binary radius of influence, $\rho_{r_{\rm infl,b}}$. The tightness of this correlation indicates that the rate at which the binary orbit shrinks during the stellar scattering phase is directly coupled to the density of the stellar reservoir in the binary loss cone~\citep{Quinlan97}; our simulation suite successfully reproduces this result, which is a promising sign that it is resolving the interactions between the black hole binary and individual star particles. It is tempting to relate $s$ or $\rho_{r_{\rm infl,b}}$ directly to the binary coalescence time, but the right panel of Figure~\ref{fig:hardening} indicates that coalescence time is driven by additional physics. Indeed, for a given hardening parameter, the coalescence time is different by a factor of 100; most of this difference can be explained by whether the secondary black hole enters in a prograde (pink) or retrograde (green) orbit with respect to the rotation of the galaxy.
Figure~\ref{fig:eccentricity} further hints that coalescence time may be best described as a manifold in a higher-dimensional feature space, with large scatter and a clear difference in coalescence time for different binary black hole orbits when plotted as a function of galaxy stellar mass (left), binary black hole mass (middle) or binary eccentricity in the gravitational wave regime (right). 

\begin{table*}
\caption{\bf Host Galaxy Properties.}
\begin{center}
\vspace{-0.5pt}
\begin{tabular}{l c c c c c c c c c c}
\hline
Galaxy & S{\'e}rsic index & $M^*_{\rm gal}$ & $\rho_{0}$ & $R_{\rm eff}$  & $v/\sigma$ & $M_{\rm NSC}$ & $R_{\rm NSC}$  & $M_{\mathrm{BH1}}$  & $r_{\rm infl}$ & $\rho_{\rm r_{infl}}$ \\
 &  & $10^9$ M$_{\odot}$ & M$_{\odot}/{\rm pc^3}$ & kpc &  & $10^7$ M$_{\odot}$ & pc & $10^6$M$_{\odot}$ & pc  & $M_{\odot}/{\rm pc^3}$ \\
\hline
NGC2639 & $3.34$ & $720$ & $1 \times 10^4$ & $5.63$ & $0.5$ & -- & -- & $1700$ & $130$ & $290$\\

NGC1374 & $3.7$ & $33.1$ & $2 \times 10^4$ & $2.34$ & $0.65$ & -- & -- & $580$ & $175$ & $2.1\times 10^4$\\

NGC6958 & $3.3$ & $36$ & $1 \times 10^3$ & $2.59$ & $0.5$ & -- & -- & $360$ & $160$ & $18$\\

MW  & $1.32$ & $9.6$ & $1 \times 10^7$ & $1.0$ & $0.6$ & $4.0$ & $4.2$ & $4.0$ & $1.2$ & $2.6 \times 10^5$\\

M32 & $1.6$ & $1.04$ & $1 \times 10^7$ & $0.18$ & $0.7$ & $1.45$ & $4.4$ & $2.5$ & $1.61$ & $6.5 \times 10^4$\\

NGC5102 & $3.0$ & $5.98$ & $2 \times 10^6$ & $1.2$ & $0.6$ & $0.71$ & $1.6$ & $0.88$ & $1.2$ & $2.0 \times 10^5$\\

NGC5206 & $2.57$ & $2.43$ & $1 \times 10^6$ & $0.99$ & $0.25$ & $0.17$ & $3.4$ & $0.47$ & $1.0$ &  $6.0 \times 10^4$\\

NGC404 & $2.5$ & $0.86$ & $2 \times 10^5$ & $0.675$ & $0.2$ & $0.34$ & $1.6$ & $0.07$ & $0.35$ & $1.0 \times 10^5$\\

NGC205 & $1.4$ & $0.97$ & $1 \times 10^6$ & $0.52$ & $0.2$ & $0.18$ & $1.3$ & $0.04$ & $0.14$ & $6.0 \times 10^5$\\






\hline
\end{tabular}\label{tab:hostgxy}

\end{center}

\begin{flushleft} 
\footnotesize{Column 1: Galaxy model; Column 2:  S{\'e}rsic index of the stellar profile; Column 3: Stellar mass of the primary galaxy in units of $10^9$  M$_\odot$, Column 4: Central stellar density in units of M$_{\odot}/{\rm pc^3}$; Column 5: Effective radius of the primary in units of kpc; Column 6: Ratio of rotational velocity over 3-d stellar velocity dispersion; Column 7: Mass of the nuclear star cluster in units of $10^7$ M$_\odot$; Column 8: Scale radius of the nuclear star cluster in units of parsecs; Column 9: Central black hole mass in units of $10^6$ M$_\odot$; Column 10: Radius of influence of the primary black hole in units of parsecs; Column 11: Stellar density at the radius of influence in units of M$_{\odot}/{\rm pc^3}$.}
\end{flushleft}
\vspace{15pt}
\end{table*}

\begin{table*}
\begin{center}
\vspace{-0.5pt}
\caption{\bf MBH Binary Evolution Runs} 
\begin{tabular}{l c c c c c c c c }
\hline
Galaxy Run & $M_{\rm BH2}$ & $M_{\rm BBH}$  & $r_{\rm infl,b}$  & $\rho_{\rm r_{infl,b}}$  & $s$ & $e_{\rm i}$ & $e_{\rm f}$ & $T_{\rm coal}$\\
 & $10^6$ M$_{\odot}$ & $10^6$ M$_{\odot}$  & pc  & $M_{\odot}/{\rm pc^3}$  & $ {\rm pc^{-1}/Myr}$ &  &  & ${\rm Gyr}$\\
\hline
NGC2639-p  & $1700$ & $3400$ & $257$ & $67$ & $0.0057$ & $0.14$ & $0.21$ & $0.742$ \\
NGC2639-p4 & $425$ & $2125$ & $178$ & $136$ & $0.012$ & $0.05$ & $0.05$ & $0.61$ \\
NGC2639-r4 & $425$ & $2125$ & $180$ & $145$ & $0.011$ & $0.92$ & $0.99$ & $0.07$ \\

NGC1374-p & $580$ & $1160$ & $431$ & $10.1$ & $0.00085$ & $0.05$ & $0.02$ & $6.75$ \\
NGC1374-r & $580$ & $1160$ & $411$ & $10.2$ & $0.00068$ & $0.82$ & $0.95$ & $1.46$ \\
NGC1374-p4 & $145$ & $725$ & $263$ & $10.4$ & $0.00265$ & $0.04$ & $0.05$ & $3.97$ \\
NGC1374-r4 & $145$ & $725$ & $257$ & $10.4$ & $0.00264$ & $0.8$ & $0.98$ & $0.37$ \\

NGC6958-p4 & $90$ & $450$ & $232$ & $10.8$ & $0.003$ & $0.06$ & $0.07$ & $4.775$ \\
NGC6958-r4 & $90$ & $450$ & $226$ & $10.4$ & $0.0029$ & $0.98$ & $0.99$ & $0.256$ \\
NGC6958-p10 & $36$ & $396$ & $190$ & $15.6$ & $0.0049$ & $0.17$ & $0.15$ & $3.96$ \\
NGC6958-r10 & $36$ & $396$ & $185$ & $15.5$ & $0.0037$ & $0.98$ & $0.985$ & $0.34$ \\

MW-p & $4.0$ & $8.0$ & $5.16$ & $3.4 \times 10^3$ & $0.78$ & $0.05$ & $0.05$ & $0.578$ \\
MW-p4 & $1.0$ & $5.0$ & $2.6$ & $4.1 \times 10^4$ & $5.13$ & $0.14$ & $0.18$ & $0.176$ \\
MW-r4 & $1.0$ & $5.0$ & $2.7$ & $4.1 \times 10^4$ & $3.3$ & $0.72$ & $0.98$ & $0.034$ \\
MW-p10 & $0.4$ & $4.4$ & $2.2$ & $7.4 \times 10^4$ & $13.2$ & $0.16$ & $0.18$ & $0.104$ \\
MW-r10 & $0.4$ & $4.4$ & $2.2$ & $7.4 \times 10^4$ & $11.5$ & $0.8$ & $0.83$ & $0.05$ \\

M32-p & $2.5$ & $5.0$ & $7.5$ & $2.89 \times 10^3$ & $0.47$ & $0.03$ & $0.02$  & $1.014$ \\
M32-p4 & $0.62$ & $3.1$ & $6.5$ & $3.68 \times 10^3$ & $1.8$ & $0.1$ & $0.08$  & $0.489$ \\
M32-r4 & $0.62$ & $3.1$ & $6.5$ & $3.71 \times 10^3$ & $1.9$ & $0.98$ & $0.91$ & $0.162$ \\
M32-p10 & $0.25$ & $2.75$ & $5.5$ & $4.78 \times 10^3$ & $2.7$ & $0.13$ & $0.14$  & $0.437$ \\
M32-r10 & $0.25$ & $2.75$ & $5.6$ & $4.7 \times 10^3$ & $2.86$ & $0.97$ & $0.90$ & $0.128$ \\

NGC5102-p & $0.88$ & $1.76$ & $1.9$ & $4.5 \times 10^4$ & $8.8$ & $0.18$ & $0.54$ & $0.176$ \\
NGC5102-r & $0.88$ & $1.76$ & $1.9$ & $4.5 \times 10^4$ & $8.3$ & $0.45$ & $0.99$ & $0.014$ \\

NGC5206 & $0.47$ & $0.94$ & $2.9$ & $2.0 \times 10^4$ & $3.73$ & $0.1$ & $0.4$ & $0.523$ \\
NGC5206-e & $0.47$ & $0.94$ & $2.9$ & $2.0 \times 10^4$ & $3.73$ & $0.1$ & $0.92$ & $0.173$ \\

NGC404 & $0.07$ & $0.14$ & $0.7$ & $1.1 \times 10^5$ & $49$ & $0.2$ & $0.5$ & $0.218$ \\
NGC205 & $0.04$ & $0.08$ & $0.25$ & $7.0 \times 10^5$ & $299$ & $0.2$ & $0.39$ & $0.074$ \\
NGC205-4 & $0.0055$ & $0.0275$ & $0.255$ & $5.3 \times 10^5$ & $223$ & $0.08$ & $0.09$ & $0.205$ \\





\hline
\end{tabular}\label{tab:resultsim1}
\begin{flushleft}
Column 1: Simulation run. The base name corresponds to the primary galaxy model as described in Table 1. The suffix after the dash, if there is one, reflects properties of the incoming black hole orbit: the letter refers to the sense of rotation of the black hole with respect to the angular momentum axis of the primary galaxy, and the number refers to the mass ratio of the primary to secondary black hole. Specifically, p is prograde, r is retrograde, no letter implies the primary was not rotating, and e corresponds to a run in which the black hole orbit was set to at a higher eccentricity shortly after hardening. Note that if no number is present, the black holes have equal mass; Column 1: Secondary black hole mass in units of $10^6$ M$_\odot$; Column 2: Total mass of the black hole binary in units of $10^6$  M$_\odot$, Column 3: Radius of influence of the black hole binary in units of parsecs; Column 4: Stellar density at the binary radius of influence in units of M$_{\odot}/{\rm pc^3}$; Column 5: Hardening rate in units of pc$^{-1}$/Myr; Column 6: eccentricity at the time of binary formation; Column 7: eccentricity in the gravitational wave regime at the end of the simulation; Column 8: Binary black hole coalescence time in units of Gyr.
\end{flushleft}
\end{center}
\vspace{15pt}
\end{table*}

\section{Methods for Finding the Fit}

Since most any regression technique will require features with similar dynamic range or unit scaling, we transformed several features before searching for our merger timescale relation. We logged all masses, densities and the coalescence time, and we centered each feature by subtracting the mean. Note that our input data only samples purely prograde or retrograde projections; we will use a separate suite with a range of orbital plane projections to test the model.

Linear regression, combined with best subset selection, appears to be the optimal approach for selecting a good predictor of the coalescence time and building a highly interpretable model from observational features. 
There are downsides to this approach: specifically, we assume that the dependent variable, coalescence time, is a linear function of some combination of features, and linear regression may only explore a limited number of models, if the number of features is quite large.
However, other methods had their downsides as well. Principal Component Analysis (PCA), for example, identifies a linear combination of features that generate a manifold along which variation in the data occurs, and its cousin, Kernel PCA, extends the PCA technique for nonlinearly related features. While we did identify a fit with PCA, it resulted in a more complex model, in terms of either having fewer interpretable principal components or requiring more essentially redundant terms. Kernel PCA, on the other hard, is best applied when we can make stronger assertions about the precise kind of nonlinearity, which we don't apriori know. More unsupervised methods, such as topological data analysis (TDA), could be promising because they would break our assumption of linearity~\citep{chazal21}. In TDA, the data are treated as a weighted graph in a high-dimensional space, and the analysis looks for clusters in the data to determine the shape of this manifold within that space. The problem is that many of the motivations for TDA were drawn from the analysis of large data sets, and with the small data set we have here, finding clusters in the data and determining accurate covariance matrices are, in practice, not robust.  If there were many more binary black hole simulations in the data set, it is possible that building a model by treating the data topologically would be the most accurate method to break our linearity assumption and approximate a nonlinear manifold that truly represents the data. We caution, though, that the result may be uninterpretable and therefore may not serve as useful a purpose as a handy scaling relation to use as a subgrid model for binary black hole merger times. After trying these fitting techniques, we found that simple linear regression best suited our needs and ultimately resulted in a highly predictive model.

\subsection{Performing the Best Subset Regression}

For subsets of size n, we tested ${8 \choose n}$ models.
For $n=$ 1, 2, 3, and 4, we tested 8, 28, 56, and 70 models, respectively. As we discuss below, by looking at the top models for each subset size, we can trace when valid subsets emerge, as well as gauge when strong multicollinearity emerges.

For each subset, we calculate the coefficient of determination, $R^2$, as a measure of how well the data fit the model. $R^2$ increases as we increase n, so we adjust $R^2$ to compare models of different sizes such that:
 
\begin{equation}
R^2_{\rm adjusted} = {1 - {\Big( {{RSS/(m-n-1)} \over {TSS/ (m-1)}} \Big)}} = {1 -(1-R^2) {{m-1} \over {m-n-1}}},
\end{equation}
where m is our number of observations, RSS is the residual sum of squares (related to variance to the model), and TSS is the total sum of squares (related to variance of the mean). This is equivalent to $R^2 = 1-{{RSS} \over {TSS}}$, except $R^2_{\rm adjusted}$ punishes each addition of a new predictor variable. 
\begin{figure*}
    \centering{
    \resizebox{0.85\hsize}{!}{\includegraphics[angle=0]{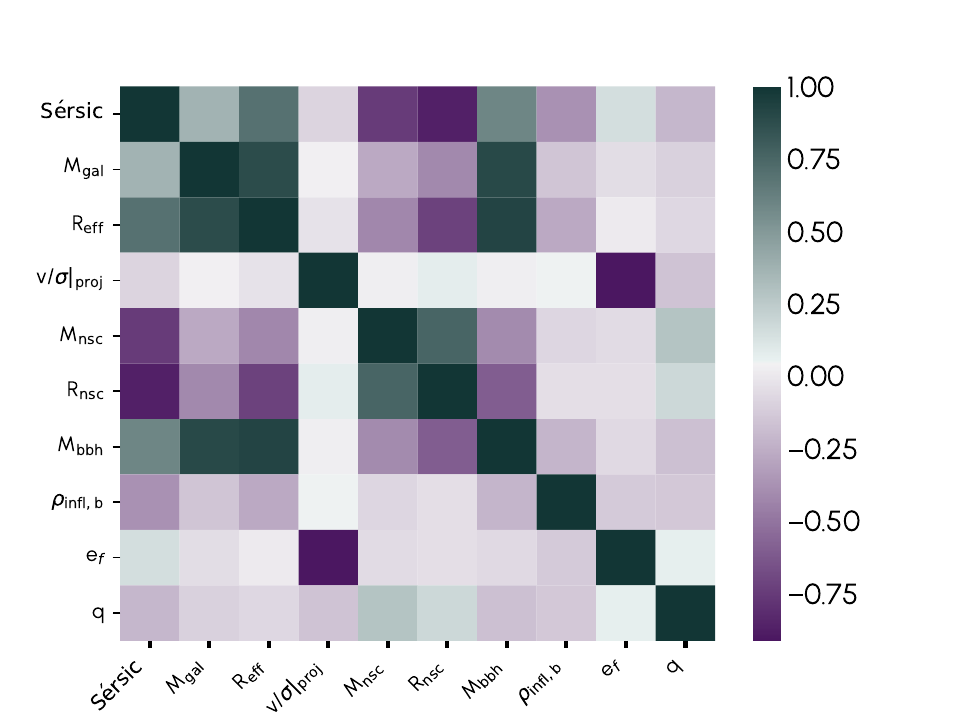}}
    }
    \caption{ Correlation matrix for selected features used in this study. The intensity of the hue represents the correlation strength, with green indicating a positive correlation and purple representing an anti-correlation. It is clear that there are very few features that are uncorrelated, which indicates that care must be taken to address multicollinearity in potential models. }
    \label{fig:correlation}
\end{figure*}

It is clear from Figure~\ref{fig:correlation} that several features are correlated with one another, and this {\it multicollinearity} can plague the robustness of any model fit. The standard measure of multicollinearity used is a variance inflation factor (VIF); the VIF for each predictor $X_i$ in a given model is: VIF$_{X_i} = 1/ (1-R^2_i)$, where $R_i$ is determined by performing regression on $X_i$ by the other predictors.
The more collinear $X_i$ is with the other predictors, the greater the VIF, and the more the model would be expected to vary if based on other random samples. Generally, ${\rm VIF} > 5$ indicates a degree of collinearity that renders a model problematic. We calculated the VIF for each model and discarded any with VIFs above this threshold. As can be seen by Figure~\ref{fig:correlation}, $\rho_{\rm infl}$ is the most colinear variable, often resulting in models that were discarded due to large VIFs.

\section{Coalescence Time Model}

For n=3, $v / \sigma|_{\rm proj}$, log$\,M_{\rm BBH}$, and log$\, \rho_{\rm infl}$ emerged as strongly predictive features and resulted in high $R_{\rm adjusted}$ models. However, the VIF between log$\,M_{\rm BBH}$, and log $\,\rho_{\rm infl}$ was 6.1, indicating a too high a degree of collinearity between these variables. The best subset of features with VIFs $< 5$ was: $v / \sigma|_{\rm proj}$, log$\,M_{\rm NSC}$, and $R_{\rm eff}$, yielding an adjusted $R^2_{\rm adjusted}=0.82$:

\begin{equation}
\log(\rm{T_{coal}}) = 9.8 - 0.21 \log( \rm{M_{NSC}})  + 0.7 (v /\sigma|_{\rm proj}) - 0.27(\rm{R_{eff}}).
\end{equation} 
\label{eq:fit} 

\noindent At n=4, $R_{\rm eff}$, $v /\sigma|_{\rm proj}$, log$\,M_{\rm NSC}$, and log$\,M_{\rm BBH}$ exhibited both a high adjusted $R^2_{\rm adjusted} = 0.86$, and VIFs=[1.9, 1.0, 4.8, 4.5], respectively:  

\begin{equation}
\log(\rm{T_{coal}}) = 8.2 + 0.19\log( \rm{M_{BBH}}) - 0.16 \log( \rm{M_{NSC}})  + 0.71 (v /\sigma|_{\rm proj}) - 0.32(\rm{R_{eff}}).
\end{equation} 
\label{eq:fit2}

\begin{figure*}
    \centering{
{\includegraphics[scale=0.45,angle=0]{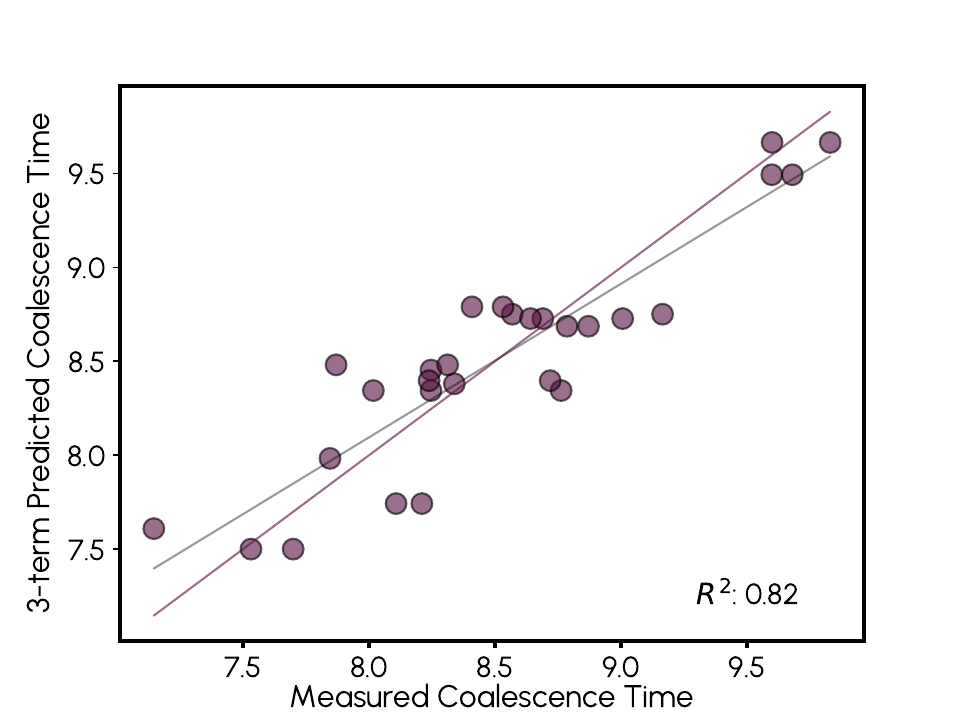}}{\includegraphics[scale=0.45,angle=0]{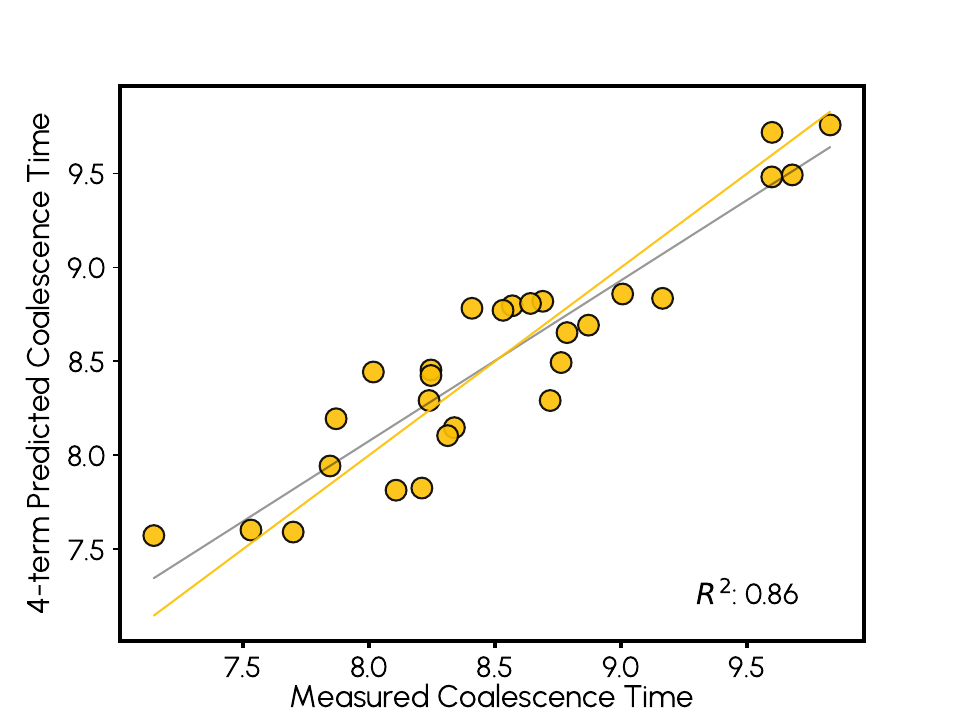}}}
    \caption{ Measured versus Predicted Coalescence Time. The colored points represent the predicted coalescence time for the 3-term model (left, see equation 4), and the 4-term model (right, see equation 5) for a given measured coalescence time, with the least-squares fit to these data plotted in a matching color (left: purple; right: yellow).  The black line represents a perfect prediction. Using a coalescence model with $R_{\rm eff}$, $v /\sigma|_{\rm proj}$, log$\,M_{\rm NSC}$, and log$\,M_{\rm BBH}$, we can reduce the variance in merger time predictions from two orders of magnitude to less than a factor of roughly 2.}
    \label{fig:predict}
\end{figure*}

\begin{figure*}
    \centering{
    \resizebox{0.95\hsize}{!}{\includegraphics[angle=0]{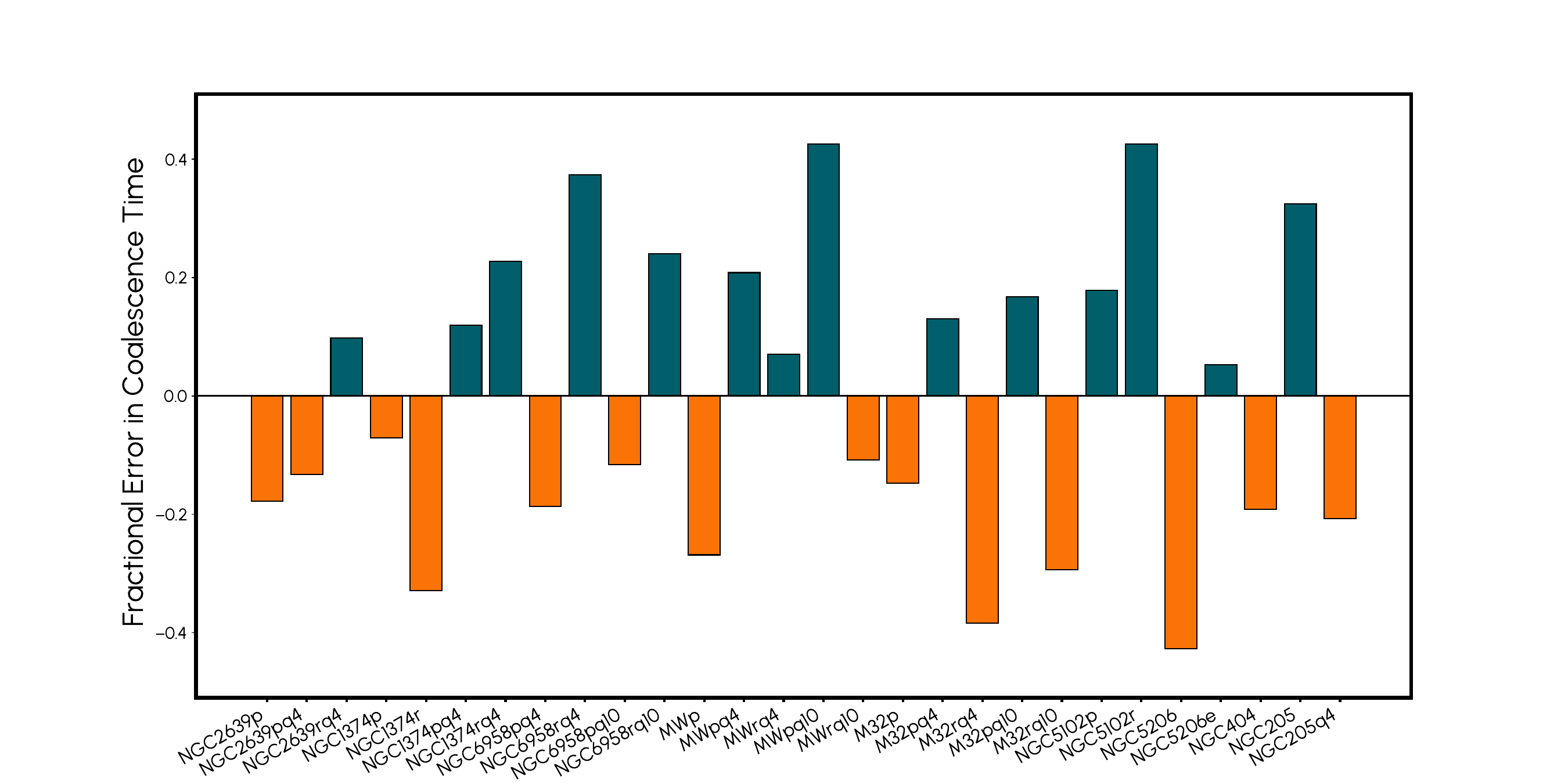}}
    }
    \caption{Fractional difference in coalescence time prediction for the galaxy runs described in Table \ref{tab:resultsim1}, using the 4-term model (equation 5). A typical coalescence time prediction will under or over estimate the true coalescence time by $ \sim 20 \%$. There is a slight hint that the predicted coalescence timescales for retrograde orbits have a larger scatter and a small bias toward overestimating the true coalescence timescale. }
    \label{fig:residual}
\end{figure*}

\noindent For most models with more than 4 terms, multicollinearity becomes a more insurmountable problem, as measured by the variance inflation factors. However, $v / \sigma|_{\rm proj}$ and q appeared most often in well-fitting models with one or more of the other features. The improvement in the adjusted $R^2_{\rm adjusted}$ by adding another term to the model was of order $0.01$ at the risk of larger multicollinearity. 
Figure~\ref{fig:predict} shows the predicted versus measured binary coalescence time for the 3 and 4-term models from equations 4 and 5, respectively. Note that for a given measured coalescence time, the range of predicted coalescence times is typically less than a factor of 2, paving the way to a more accurate treatment of binary black hole dynamics in subgrid models.   Figure~\ref{fig:residual} shows the fractional error in the predicted binary coalescence timescale compared to the actual timescale for the 4-term model. Half of the predictions are accurate to within 20\% of the true coalescence time, and about a quarter of the models have errors of 40\%. Most of these larger outliers have retrograde configurations, which hints at a slight bias in the model; future simulations are needed to flesh this out.

\begin{figure*}
    \centering{
    \resizebox{0.95\hsize}{!}{\includegraphics[angle=0]{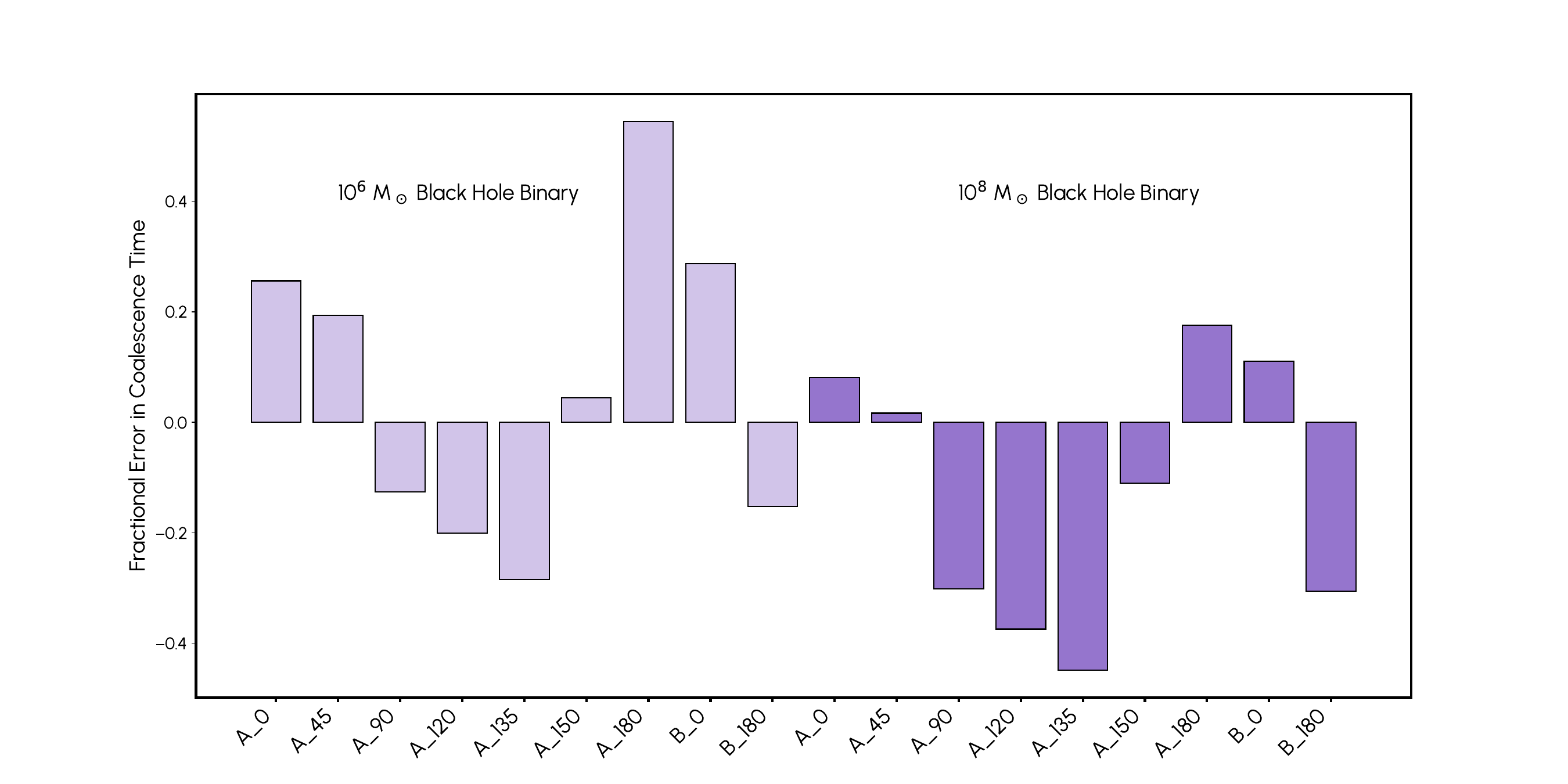}}
    }
    \caption{Fractional error in coalescence time for our test suite. The simulations were taken from \citet{Mirza+17}, where the A models explore the effect of binary orbital inclination with respect to the galaxy plane, and the B models investigate the effect of counter-rotating galaxy cores. The subscript corresponds to $\Theta$, the angle between the angular momentum axis of the binary black hole orbit and the angular momentum axis of the galaxy, where $0^{\circ} < \Theta < 90^{\circ}$ corresponds to corotation of the binary with the stellar material of the galaxy and $90^{\circ} < \Theta < 180^{\circ}$ corresponds to counterrotation.}
    \label{fig:test}
\end{figure*}

\subsection{Testing the model}

Testing the model required another set of data, which was drawn from a suite of simulations that explored coalescence time as a function of the orientation of the binary black hole orbital plane with respect to the rotation of the galaxy model~\citep{Mirza+17}, where the galaxy model has an otherwise fixed shape and structure~\citep[e.g.][]{KHB15}. Since $v / \sigma|_{\rm proj}$ and black hole masses were highly significant predictors in our best subset models, and since there was a missing range of values for $v / \sigma|_{\rm proj}$ in our original dataset, choosing a sample which only varies those two predictors provides a good, but basic, test for our model. Specifically, there were 9 $v / \sigma|_{\rm proj}$ values, as the orbital inclination ${\rm Cos }\Theta$ varied, and 2 black hole masses, for a total of 18 models (see ~\citep{Mirza+17} for more details). Notably, the hosts in this test suite did not include nuclear star clusters, the mass of which is a significant predictor in both models. Figure~\ref{fig:test} shows the fractional error in the predicted binary coalescence time using the 4-term model compared to the measured coalescence time in ~\citet{Mirza+17}. Our model is slightly more likely to overestimate the binary black hole scattering timescale for lower mass black holes; perhaps this is due to the fact that the hosts for these galaxies typically harbor nuclear star clusters, which are missing from the test suite.   Nonetheless, we find a moderately accurate linear fit for the clearly nonlinear relationship between $v / \sigma|_{\rm proj}$ and coalescence time when holding all other variables fixed, with most coalescence time predictions being accurate to within 30$\%$.

\section{Caveats, Implications, and Future Work}

We identified a handy relation to predict binary black hole coalescence timescales for the stellar scattering phase, based on observable properties of the host galaxy and the incoming black hole orbit. This relation holds for a wide range of black hole masses, from intermediate to supermassive, and a variety of host galaxy morphologies, including dwarf galaxies, disk galaxies, and giant ellipticals. We find that the stellar scattering phase spans between roughly 10 Myr to a few Gyr; the binary black hole mass, presence of a nuclear star cluster, degree of rotation and orientation with respect to the incoming black hole orbit, and galaxy size are key ingredients in determining this timescale. 

Although our relation is successful at predicting coalescence timescales in this stage, there are a number of caveats to keep in mind, both procedural and astrophysical. For example, since galaxy characteristics themselves are tightly coupled (the Fundamental Plane is a perfect example of this), it is difficult to avoid variables that are highly correlated. However, simple regression is based on the assumption that features are independent, so that the true regression coefficients are independent as well, and when this assumption fails, the resulting model may not be robust. To combat this, we combined or dropped highly correlated variables. Even so, the resulting set of features, M$_{\rm BBH}$, M$_{\rm NSC}$, R$_{\rm eff}$, and $v/\sigma|_{\rm proj}$ are not all strictly independent. In addition, because these simulations are so computationally expensive, we are formally dealing with a sparse data set; a good rule of thumb suggests that there be at least 10 data points per predictor, though this number could be significantly higher depending on the complexity of the problem. This is largely because clusters in the data can lead to misleading correlations and, consequently, misleading results. With fewer than 30 runs and nearly twenty features, there is not enough data to adequately span all features. Here, the correlations between features in a galaxy population are actually helpful; by dropping sets of features in a prospective model that were highly correlated, we restricted the number of features in a model to 3-4. Fortunately, since regression is essentially fitting a hyperplane to the data, the hyperplane should still locally fit the data along the regression lines as long as the underlying correlations between features are real. 

 Understandably, it will be difficult to test any model until there are more data that better fill the feature space. We tested our model with a suite of simulations that vary our more heavily weighted features, black hole mass and $v/\sigma|_{\rm proj}$, holding other features constant. We found that our relation was a good predictor of coalescence time in the test suite, but this is only a single test that verified that our relation is not deformed by existing clusters in the data along these two features. Nonetheless, given this single test, it remains plausible that we have found a local linear fit for the coalescence time in the scattering phase as a function of M$_{\rm BBH}$, M$_{\rm NSC}$, R$_{\rm eff}$, and $v/\sigma|_{\rm proj}$ that is accurate to within a factor of two. This is two orders of magnitude better than prior predictions that rely merely on stellar density.

Though the galaxy models in our suite are based on observed galaxies, they do not represent all galaxy types, nor do they contain all relevant physics that may influence MBH binary coalescence once the binary is bound. First and foremost, gas is ever present in galactic nuclei and has long been thought to accelerate the binary inspiral through gravitational torques~\citep[e.g.][]{artymowicz+1994, artymowicz+1996, gould2002, armitage2002, escala+05, cuadra+2009, goicovic+2018, tiede+2020, heath2020}. This picture becomes more complex when considering more realistic physics such as star formation and feedback from an Active Galactic Nucleus (AGN). Most cosmological scale simulations that include gas effects assume a binary will be universally driven inwards~\citep{haiman+2009, kelley+2017a, kelley+2017b, volonteri+2020, volonteri+2022}. AGN feedback can evacuate a central cavity \citep{Khan2025} and suppress accretion-driven torques leading to stages of slowed or even stalled binary evolution~\citep{delValle+2018, dEtigny+2024}. A number of recent studies find that under certain conditions binaries may widen~\citep{miranda+2017, tang+2017, moody+2019, munoz+2019, munoz+2020, duffell+2020, duffell+2024}. The emergence of outward migration is highly sensitive to numerical simulation assumptions and disk properties, and the physical relevance of the regimes in which these effects manifest is still an active area of investigation~\citep[e.g.][]{dOrazio2021, dittmann2022, dittmann2024, franchini+2022, penzlin+2022, sudarshan+2022, siwek+2023}. These highly non-linear systems highlight the need for further study to characterize the effects of stars and gas in binary hardening. Observations of coalescing MBH binaries with gravitational wave detectors like LISA can offer insight by constraining the environmental signatures imparted on the gravitational waveforms~\citep{derdzinski+2019, derdzinski+2021, garg+2022, cole+2023, spadaro+2025}.
 
There are also at least two types of galactic nuclei that have been excluded from our simulation suite. About 20-40$\%$ of local optically-selected disk galaxies contain bars~\citep[e.g.][]{Aguerri+09, Masters+11, Euclid+25}, with this fraction rising to 70$\%$ or more if weak bars are included, or if the galaxy is viewed in the infrared~\citep[e.g.][]{Eskridge+20, Sheth+08, Menendez+07}. 
On a larger scale, infalling MBHs can have difficulty making their way to the galactic center as they are swept up by bar-induced resonances~\citep[e.g.][]{Bortolas+22}.  In addition, 
we exclude the lowest density {\it non-nucleated} dwarf galaxies ($M_{\text{gal}} \lesssim 10^7 M_{\odot}$). The existence of central massive or intermediate mass black holes in these extreme systems is highly unconstrained due to observational limitations, but if local scaling relations follow, these systems may be expected to host some of the lightest MBHs~\citep{Reines15, Greene20}. Direct $N$-body simulations of these low density dwarf galaxies show that MBH binaries struggle to merge within a Hubble time due to the lack of stellar material for scattering~\citep{khan2024}. If these dwarfs do indeed host MBHs and have grown through hierarchical merging, they may host a population of low-mass MBH binaries that have stalled indefinitely, an archival history of the assembly of the galaxy.

Finally, all our initial galaxy models were constructed in equilibrium, which neglects the rich variety of gravitational perturbations that shape a galaxy within a cosmological context. Over a typical stellar scattering timescale of $\mathcal{O}(10^8)$ years, treating the host galaxy as quasi-isolated potential is arguably well motivated, but this assumption is less realistic on larger timescales as the galaxy itself transforms in mass, morphology, and kinematics -- or at high redshift when the host galaxy is actively assembling~\citep{Dicesare23, Harikane23, Perez23}. In addition, the secondary black hole would presumably be shrouded by the remnants of its own galactic center, and this extra mass would accelerate the coalescence in this stage both by perturbing the potential and by bringing in a fresh supply of stars to the loss cone~\citep{khan18b, ogi19, muk2025}.

Despite these caveats, the fact that the stellar scattering phase can be estimated so well with bulk characteristics of the galaxy host and the initial orientation of the incoming black hole makes it a boon for observational and theory work alike. It would be particularly useful to deploy these relations to better refine predictions of the occupation fraction of supermassive black hole binaries, or to calculate MBH merger rates for LISA and NANOGrav from cosmological $N$-body simulations. We explore these uses in an upcoming paper~\citep{Roth25}.
 
    \section*{Acknowledgments}
    
This project started during COVID and experienced nearly as many delays as a black hole does in reaching a galactic center. KHB appreciates her co-authors' patience with the coalescence timescale of this manuscript and thanks them for encouragement and helpful suggestions along the way. This work was run on ACCRE at Vanderbilt University and was supported by NASA-NNX08AG74G, NSF-2125764, and NSF-2319441. FMK and AVM acknowledge the support by Tamkeen under the NYU Abu Dhabi Research Institute grant CASS.
    
	\section*{Data Availability Statement}
	
	The data underlying this article will be shared on reasonable request to the corresponding author.
	
	

	\bibliographystyle{mnras}
	\bibliography{ms}

	\bsp	
	\label{lastpage}
\end{document}